\documentclass[a4paper,prb,twocolumn]{revtex4}
\usepackage[dvips]{graphics,color}
\usepackage[pdftex]{graphicx}
\usepackage{amsmath}
\usepackage{amsfonts}
\usepackage{amssymb}
\usepackage{amstext}
\usepackage[sort&compress]{natbib}

\newcommand{\id}{\mathbb{I}}

\newcommand{\beq}{\begin{equation}}
\newcommand{\eeq}{\end{equation}}
\newcommand{\beqa}{\begin{eqnarray}}
\newcommand{\eeqa}{\end{eqnarray}}
\newcommand{\beqar}{\begin{eqnarray*}}
\newcommand{\eeqar}{\end{eqnarray*}}
\newcommand{\tr}{{\rm tr}}

\newcommand{\bra}[1]{\langle #1 |}
\newcommand{\ket}[1]{| #1 \rangle}

\begin{document}

\title{Entanglement Theory and the Second Law of Thermodynamics}

\author{Fernando G.S.L. Brand\~ao$^{1,2}$ \& Martin B. Plenio$^{1,2}$}

\affiliation{QOLS, Blackett Lab, Imperial College London,
Prince Consort Road, London SW7 2BW, UK}
\affiliation{Institute for Mathematical Sciences, Imperial College
London, 53 Exhibition Road, London SW7 2PG, UK}

\date{\today}

\begin{abstract}
Entanglement is central both to the foundations of quantum 
theory and, as a novel resource, to quantum information 
science. The theory of entanglement establishes basic laws, 
such as the non-increase of entanglement under local operations, 
that govern its manipulation and aims to draw from them formal 
analogies to the second law of thermodynamics. However, while 
in the second law the entropy uniquely determines whether a 
state is adiabatically accessible from another, the manipulation 
of entanglement under local operations exhibits a fundamental 
irreversibility which prevents the existence of such an order.

Here we show that a reversible theory of entanglement and a 
rigorous relationship with thermodynamics may be established 
when one considers all non-entangling transformations. The 
role of the entropy in the second law is taken by the asymptotic 
relative entropy of entanglement in the basic law of entanglement. 
We show the usefulness of this new approach to general resource 
theories and to quantum information theory. 
\end{abstract}
\maketitle

Thermodynamics is arguably one of the most fundamental and 
generally applicable theories of Nature whose foundations
have remained intact despite the emergence of quantum mechanics,
relativity and other physical laws\cite{Thermo}. It was 
initially understood to describe the physics of large systems 
in equilibrium, determining their bulk properties by a very simple 
set of rules of universal character. This was reflected in 
the formulation of the defining axiom of thermodynamics, the 
second law, by Clausius, Kelvin and Planck in terms of quasi-static 
processes and heat exchange. However, the apparently universal 
applicability of the theory suggested a deeper mathematical 
and structural foundation. Indeed, there is a long history of
examinations of the foundations underlying the second law. 
Of particular interest in the present context is the work
of Giles\cite{Giles} and notably Lieb and Yngvason\cite{Lieb} 
stating that there exists a complete order for equilibrium 
thermodynamical states that determines which state transformations 
are possible by means of an adiabatic process. From 
simple, abstract, axioms they were able to show that this 
order is uniquely determined by an \textit{entropy} function 
$S$: given two equilibrium states $A$ and $B$, $A$ can be 
converted by an adiabatic process into $B$ if, and only if, 
$S(A) \leq S(B)$. As pointed out by Lieb and Yngvason\cite{Lieb2},
it is a strength of this abstract approach that it allows 
to uncover a thermodynamical structure in settings that 
may at first appear unrelated. 

One such possible setting is the theory of entanglement 
\cite{Plenio1,Horodecki0} in quantum information science
\cite{Bennett00}. Although the importance of entanglement 
for the foundations of quantum mechanics has been noticed 
early\cite{Einstein, Schrod}, it was only over the last 
decade or so that the resource character of quantum 
correlations was recognized. A paradigmatic scenario in 
this respect is the one in which two distant parties want to 
exchange quantum information, but are restricted to act 
locally on their quantum systems and communicate classically. 
It is known that quantum state teleportation \cite{Bennett BCJPW 92} 
overcomes the limitations caused by the restriction to local
operations through the use of quantum mechanical correlations, 
entanglement. Entanglement theory is then concerned 
with the systematic exploration of entanglement as a resource 
in a quantitative manner.

Possible connections between entanglement theory and 
thermodynamics were noted when it was found that for 
bipartite pure states a very similar situation to the 
second law holds in the asymptotic limit of an arbitrarily 
large number of identical copies of the state. Then, 
given two bipartite pure states $\ket{\psi_{AB}}$ and 
$\ket{\phi_{AB}}$, the former can be converted into the 
latter by local operations and classical communication 
(LOCC) if, and only if, $E(\ket{\psi_{AB}}) \geq E(\ket{\phi_{AB}})$, 
where $E$ is the entropy of entanglement, given by the von 
Neumann entropy of the reduced density matrix\cite{Bennett1}. 

However, for mixed entangled states there are bound
entangled states \cite{Horodecki1} that contain quantum correlations and 
thus require a non-zero rate of pure state entanglement for their creation by LOCC, 
but from which no pure state entanglement can be extracted 
at all\cite{Horodecki1, Vidal0, Horodecki22}. As a consequence 
no unique measure of entanglement exists in the general case 
and no unambiguous and rigorous direct connection to the second law
appears possible despite various interesting 
attempts\cite{Popescu1,Horodecki4,Plenio2,Plenio4,HOH02}. 

{\em Concepts and main results --}
In this work we identify the correct counterpart in 
entanglement theory of an adiabatic process. This allows
us to establish a theorem completely analogous to the Lieb and Yngvason formulation 
of the second law of thermodynamics\cite{Lieb2} for entanglement manipulation.
Similar considerations may also be applied to resources theories in general, including theories 
quantifying the non-classicality or the non-Gaussian 
character of quantum states. In the following we explain the basic relevant 
concepts, technical tools and the main result together
with conclusions that we may draw from it. 

%An outline 
%of the technical details of the proof is presented in 
%the Methods section. 

{\em Entangled states --} We begin by delineating mathematically 
the boundary between classically and quantum correlated states 
for systems that consist of two parties each holding an arbitrary number of particles. A quantum 
state of such a system is described by a density operator 
$\rho$ acting on ${\cal H}_A \otimes {\cal H}_B$, where 
${\cal H}_{A/B}$ are finite dimensional Hilbert spaces. This 
quantum state contains only classical correlations, and is 
called separable\cite{Werner1}, if there exist local density 
operators $\rho^{A/B}_j$ acting on ${\cal H}_{A/B}$ and a 
probability distribution $\{ p_j \}$ such that 
\begin{equation} 
        \rho = \sum_{j} p_j \rho^A_j \otimes \rho^B_j. 
        \label{sep}
\end{equation}
Operationally, the set of separable states ${\cal S}$ is formed by all
states that may be created from a pure product state 
$|0\rangle\langle 0|\otimes |0\rangle\langle 0|$, i.e. 
an uncorrelated state, by means of LOCC. If $\rho$ cannot be 
written as in eq. (\ref{sep}) we say it is entangled and its
generation requires, in addition to LOCC, an exchange of quantum particles (quantum communication) or a
supply of pre-existing pure entangled states that are consumed
in the process. A central state in this context is the two 
qubit maximally entangled state,
\begin{equation}  
        \phi^+ := \frac{1}{2} 
        \sum_{i,j=1}^2 \ket{i, i}\bra{j, j},
\end{equation}
which defines the unit of entanglement\cite{Plenio1}. 

The observation that LOCC alone does not create entanglement
has been taken further to formulate as the basic law of 
entanglement manipulation that entanglement cannot 
be increased by LOCC. A principle that is similar in spirit
to the second law of thermodynamics.

\begin{figure}[b]
\includegraphics[width=8.8cm]{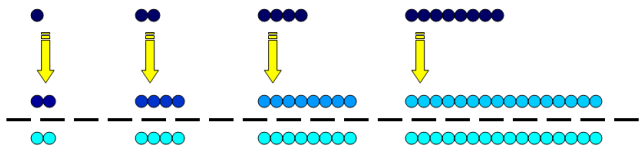}
\caption{\label{fig1} \textbf{Asymptotically entanglement conversion: } 
The dark blue circles in the top row represent
the initial state while the light blue coloured
particles in the bottom row are the target state.
The particles in the second row are the result
of a quantum operation and approximate the third 
row increasingly well with increasing number of 
particles as indicated by the colour. Asymptotically 
the approximation becomes perfect.}
\end{figure}

{\em Asymptotic entanglement conversion --} Our 
results are concerned with the asymptotic limit 
of a large number of identical copies of 
quantum states. We consider this scenario as we seek to 
indentify a total order for entanglement manipulation. 
For a finite number of copies, it is known that no such an 
order can exist, a fact that resembles the necessity of the 
thermodynamical limit in the context of the second law. 

Here it is also natural to consider transformations 
between states that may be approximate for any finite number 
of systems and are only required to become exact in the 
asymptotic limit (see fig. \ref{fig1}).
To describe this limiting process rigorously the well 
established trace distance $D(\rho, \sigma) = 
|| \rho - \sigma ||_1$ between quantum states is used. 
For two states that are close in trace distance the 
expectation value of any bounded physical observable 
will also be close when measured on the two states. 
Thus the trace distance quantifies how similar
two states behave in physical experiments and is thus a
sensible measure of distance between quantum states.

We say that a state $\rho$ can be asymptotically  converted into another 
state $\sigma$ by operations 
of a given class if there is a sequence of quantum 
operations $\{ \Psi_n \}$ within such a class acting 
on $n$ copies of the first state such that 
$\lim_{n \rightarrow \infty} D(\Psi_n(\rho^{\otimes n}), \sigma^{\otimes n}) = 0$. 

\begin{figure}[b]
\centering
\includegraphics[scale=0.3]{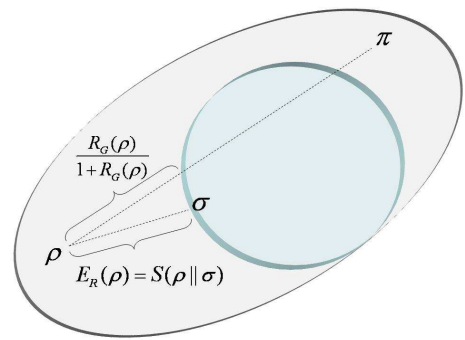}
\caption{ \textbf{Relative entropy of entanglement 
and global robustness of entanglement.} Given an entangled 
state $\rho$, one can ask how distinguishable it is 
from a separable state $\sigma$. 
A good figure of merit is the quantum relative 
entropy $S(\rho || \sigma)$ (see fig. 4). Following 
the intuition that the more entangled the more 
distinguishable a state is from a classically 
correlated one, we can then quantify the entanglement 
of $\rho$ by the minimum of $S(\rho || \sigma)$ over 
all separable states $\sigma$, finding the measure 
known as relative entropy of entanglement $E_R$\cite{Plenio3}. 
Another approach to the quantification of entanglement 
is to look at the robustness of the quantum correlations 
contained in a state to global noise, from a geometrical 
point of view. Given an entangled state $\rho$, the global robustness of entanglement 
$R_G$ is the minimal amount of mixing $s$ of $\rho$ with 
another arbitrary state $\pi$ in 
order to turn the mixture $\frac{1}{1 + s}(\rho + s\pi)$ 
into a separable state\cite{Vidal1,Nielsen1}.}
\end{figure}

{\em Measures of Entanglement --} 
To formulate the main result we need two measures 
that will be used to quantify entanglement. It will be
satisfying that these two seemingly distinct approaches
will turn out to be equivalent in the asymptotic limit, 
as a consequence of the technical proof of our main 
theorem. Firstly, we consider the relative entropy of 
entanglement\cite{Plenio3,Plenio4} 
\begin{equation}
        E_R(\rho) = \min_{\sigma \in {\cal S}} S(\rho || \sigma),
\end{equation}
where $S(\rho || \sigma) = tr\{\rho\log\rho - \rho\log\sigma\}$ 
is the quantum relative entropy. and ${\cal S}$ is the set of separable states. 
Furthermore, we consider the global robustness 
of entanglement\cite{Vidal1,Nielsen1} which is defined as 
\begin{equation}
        R_G(\rho) = \min_{s\in \mathbb{R}} 
        \left( s : \exists \hspace{0.05 cm} \sigma \hspace{0.2 cm} 
        s.t. \hspace{0.2 cm} \frac{\rho + s\sigma}{1 + s} 
        \in {\cal S} \right). 
\end{equation}
Both are meaningful entanglement quantifiers whose physical
motivations are outlined in figure 2. They will play a crucial 
role in the proof of our main theorem.

As we are concerned with the entanglement properties 
in the asymptotic limit we should not consider entanglement 
measures at the single copy level, but rather their 
asymptotic, or regularized, counterpart. In the case 
of $E_R$, the relevant quantity to consider is actually 
the regularized relative entropy of entanglement, 
given by 
\begin{equation} 
     E_R^{\infty}(\rho) = \lim_{n \rightarrow \infty} 
        \frac{E_R(\rho^{\otimes n})}{n}.
        \label{regularized}
\end{equation}
This will turn out to be the central quantity in this work
as it will emerge as the unique entanglement quantifier.

{\em Operations --} The correct choice of the set of 
operations employed is crucial for establishing reversibility in entanglement 
manipulation. To motivate this choice it is instructive to note 
that in the context of the second law it follows both 
from the approach of Giles\cite{Giles} as well as Lieb and 
Yngvason\cite{Lieb} that the class formed by all 
adiabatic processes is the largest class of operations 
which cannot decrease the entropy of an isolated 
equilibrium thermodynamical system. 

Following such an inside we now identify the largest set of quantum operations that obeys 
the basic law of the non-increase of entanglement. 
Then one might expect to achieve reversibility in entanglement 
manipulation and thus a full analogy to the second law of thermodynamics. While 
operationally well motivated, the set of LOCC operations 
is not such a class. 

The following choice of quantum operations represents 
a key insight that is crucial for establishing
reversibility in entanglement theory. 
As we are concerned with asymptotic 
entanglement manipulation, it is physically natural and also convenient 
for mathematical reasons to define the set of asymptotically 
non-entangling operations, composed of sequences of operations that for a finite number of copies $n$ may generate 
a small amount $\epsilon_n$ of entanglement which vanishes asymptotically, $\lim_{n \rightarrow \infty}\epsilon_n = 0$. 
It is important to note that we do not simply require that the 
entanglement per copy vanishes but actually the total 
amount  of entanglement. More precisely, we call a quantum 
operation $\Omega$ a $\epsilon$-non-entangling operation if 
$R_G(\Omega(\sigma)) \leq \epsilon$ for every separable 
state $\sigma$. Thus the entanglement that is generated 
by the map is not robust to a small perturbation by $\epsilon$, which 
in particular implies that the state is $\epsilon$-indistinguishable from 
a separable state in trace norm. We then call a sequence of quantum operations $\{\Psi_n\}$ asymptotically 
non-entangling when each $\Psi_n$ is an $\epsilon_n$-non-entangling 
operation and $\lim_{n\rightarrow \infty}\epsilon_n = 0$. 
This class is the largest class of quantum operations which cannot create 
entanglement in the limit of asymptotically many copies. 
Moreover, for the two measures we consider 
in this work, namely the relative entropy of entanglement 
and the global robustness of entanglement, it is also the 
largest class that asymptotically satisfies the law 
of the non-increase of entanglement. 

To demonstrate reversibility of entanglement manipulation
under the class just introduced we define two quantities 
and prove that they are the same. The \textit{entanglement cost} under asymptotically 
non-entangling maps $E_C(\rho)$ is defined as the infimum 
of $\lim_{n\rightarrow \infty}\frac{k_n}{n}$ over all the 
sequences of quantum maps $\{ \Psi_n \}$ and integers $\{k_n \}$, 
such that (i) each $\Psi_n$ is a $\epsilon_n$-non-entangling map, (ii) 
$\lim_{n\rightarrow\infty}\epsilon_n = 0$, and (iii) 
$\lim_{n\rightarrow\infty}D(\Psi_n((\phi^+)^{\otimes k_n}), 
\rho^{\otimes n}) = 0$. Note that the map might add or take out
particles from the system, so its input and output dimensions might differ. 

Conversely, the \textit{distillable 
entanglement} under asymptotically non-entangling maps 
$E_D(\rho)$ is given by the supremum of $\lim_{n\rightarrow \infty}\frac{k_n}{n}$ 
over all the sequences $\{ \Psi_n \}$ and $\{ k_n \}$ of quantum maps and integers 
satisfying conditions (i) and (ii) as in the paragraph above, but with the third requirement replaced by 
$\lim_{n\rightarrow\infty} D(\Psi_n(\rho^{\otimes n}), 
(\phi^+)^{\otimes k_n}) = 0$.

{\em The main theorem --}
Identifying the class of asymptotically non-entangling 
operations is of course not sufficient for our goals as it is not self-evident
that this class will be sufficiently powerful to permit 
the existence of reversible interconversion of entangled 
resources. It is the \textit{main result} of our work 
formulated in theorem I below to demonstrate that actually
such a class does indeed lead to reversible transformations.

\begin{figure}
  \centering
  \includegraphics[scale=0.35]{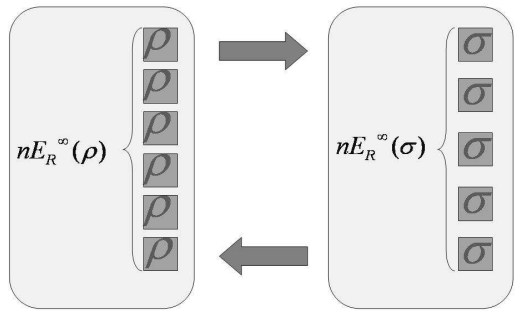}
  \caption{\textbf{Reversibility of entanglement manipulation.} 
Entanglement under asymptotically non-entangling operations is a 
fungible resource: any two entangled states can be reversibly 
interconverted in the asymptotic limit ($n \rightarrow \infty$), 
as long as the ratio of copies of each of them matches the 
ratio of the respective regularized relative entropies of 
entanglement.}
\end{figure}
 
%asymptotically non-entangling operations are indeed sufficient 
%to ensure the existence of reversible entanglement manipulation 
%in the asymptotic limit and a unique entanglement measure, the 
%asymptotic relative entropy of entanglement\cite{Plenio4,Plenio3,Plenio1}.
%Our main technical result can be rigorously stated as\\
\vspace{0.1 cm}
\hspace{- 0.3 cm}{\bf Theorem I --} \textit{For every bipartite state} $\rho$, 
\begin{equation} \label{main}
        E_C(\rho) = E_D(\rho) = E_R^{\infty}(\rho).
\end{equation}

As shown in figure 3, in this framework entanglement can be interconverted 
reversibly. The situation is formally analogous to the 
Lieb and Yngvason formulation of the second law of thermodynamics. 

We have also identified $E_R^{\infty}$ as the natural counterpart to the entropy function 
in the second law. The relative entropy measures the 
distinguishability of two quantum states (see figures 2 and 4), an observation that 
plays an important role in the technical proof. Therefore, in our framework, 
the amount of entanglement of a given state is completely 
determined by how distinguishable it is from a state 
that only contains classical correlations.    

An outline of the proof of theorem I is given in the 
Methods section. We use techniques from quantum hypothesis 
testing\cite{Petz1, stein} (see figure 4), which we generalize to 
tackle also non-identical-and-idependently-distributed instances, 
together with ideas from convex 
optimization and duality, applied to the cone of separable 
states, recent results on the characterization of symmetric states\cite{Renner}, and several properties of entanglement measures that 
have been established in recent years\cite{Plenio1, Horodecki0}.

{\em Implications and Applications --} 
Theorem I has further consequences and implications that 
we would like to explore in the remainder of this work.

{\em Entanglement Theory:} Firstly, it sheds some light 
on the origin of the irreversibility in the LOCC setting and on
the associated phenomenon of bound entanglement\cite{Horodecki1}.  
Indeed, our result suggests that such an irreversibility could
have its roots on the assymmetry of operations that can be locally implemented 
without any entanglement (LOCC) and the operations that do not generate any 
entanglement. Moreover, it also represents 
a conter-argument to linking irreversibility to the loss of classical 
information when going from pure to mixed states. 

Secondly, the uniqueness theorem we established for mixed-state 
entanglement measures is an important new conceptual tool 
in entanglement theory. Indeed, we have seen already that the 
insight of considering entanglement manipulation under asymptotic 
non-entangling maps have allowed for the identification of 
the global robustness and the relative entropy of entanglement 
in the asymptotic regime, two important measures of 
entanglement that were believed to be unrelated. 
In fact, theorem I implies that any measure which is (i) asymptotically continuous\cite{Plenio1,Horodecki0}, (ii) monotonic under 
$\epsilon$-non-entangling maps in the limit $\epsilon \rightarrow 0$, and (iii) normalized, having a value of $n$ for $\phi_+^{\otimes n}$, regularizes to $E_R^{\infty}$. From this result it is possible to identify new measures that are useful both as sharp bounds for the LOCC distillable 
entanglement and also as a first step towards the derivation 
of a single-letter expression for the regularized 
relative entropy of entanglement, i.e. a formula without needing the limit 
of a infinite number of copies of the state as in eq. (\ref{regularized}). 

Even more complicated than bipartite mixed state entanglement 
manipulation is the case of of multipartite entanglement, where already for 
pure states little is known about entanglement conversion under 
LOCC\cite{Plenio1,Horodecki0}. Our approach can be readily 
generalized to the multipartite setting, leading again to an 
unique measure of entanglement, the regularized
relative entropy of entanglement with respect to the set of
fully separable states, completely specifying states 
transformations under asymptotically non-entangling operations.

{\em General resource theories:} The approach of considering the manipulation of a given resource 
under the largest class of operations that cannot create it is appealing and can 
also be applied to other resource theories.

One example is the study of the non-classicality of quantum 
states of continuous variable systems describing e.g. 
modes of the electromagnetic field, nano-electro-mechanical
oscillators or quantum fluctuations of large atomic ensembles. Coherent states and convex combinations thereof are usually 
regarded as classical, whereas any other state is said to be 
non-classical. We can define the set of \textit{classical 
operations} as the set of operations that maps every classical state 
to another classical state. Non-classical states can then be seen as a
resource to implement general operations using only classical ones.  
The study of the manipulation of non-classical states by such operations 
emerges as a natural analog to entanglement theory and indeed basic features 
such as the non-increase of the degree of non-classicality under 
classical operations can be recovered. Both the relative entropy 
of entanglement and the robustness of entanglement have direct 
counterparts for the quantification of non-classicality, obtained 
by replacing in their definition the set of separable states 
set with the set of classical states. Hence we can also define 
the class of asymptotically classical maps as we did 
before for entanglement. 

A paradigmatic example of a non-classical 
state is the so-called cat state, given by $\ket{\phi_{\alpha}} 
:= (2(1 - e^{-|2\alpha|^2}))^{-1/2}(\ket{\alpha} - \ket{- \alpha})$, 
where $\ket{\alpha}$ is a coherent state with mean number of 
photons $|\alpha|^2$. We define in analogy to entanglement 
theory the cost of forming, with classical operations, 
a state out of many copies of $\ket{\phi_{\alpha}}$ and the distillation
rate of obtaining it from many copies of a given non-classical state. Interestingly, 
we can employ the same techniques of the proof of theorem I to show that 
the cost function in the case where $|\alpha| \rightarrow \infty$ 
converges to the asymptotic global robustness of non-classicality, 
defined in analogy to eq. (\ref{globalr}). This is result is 
remarkable because, in contrast to entanglement theory, it is not 
a priori clear that we could define a unit of non-classicality 
from which any other non-classical state could be formed. It is possible not only 
to identify the cat state of a very large number of 
excitations as such pure form of non-classicality but also provide 
a closed form formula for the cost under asymptotically classical maps.

Another example of such a resource theory concerns the Gaussian 
character of quantum states. This is of relevance in 
continuous variable implementations of quantum information 
processing in the optical regime, where Gaussian states are
most easily accessible \cite{EisertP03}. Results analogous
to the above can be obtained for the non-Gaussianity of 
quantum states as well, where now any quantum state that cannot be 
written as a convex combination of states with a Gaussian Wigner function is treated 
as a resource. 

The paradigm that we put forward could also 
potentially be helpful in the sudy of non-locality of 
quantum states, of super-selections rules \cite{GourS07}
and even beyond the real of quantum physics, in the study of 
secret correlations of tripartite probability distributions, 
in cryptography \cite{maurerwolf}. 

{\em Entanglement and Thermodynamics:} To conclude we would 
like to point out an intriguing possibility raised by our 
findings. Whereas Giles\cite{Giles} as well as Lieb and 
Yngvason\cite{Lieb,Lieb2} derived the total order for 
thermodynamical states from a set of postulates that intuitively 
should hold for thermodynamics, we have proved the existence 
of a total order for entangled states from the axioms 
of quantum theory. From theorem I we can do the opposite and 
try to derive the axioms proposed by Lieb and Yngvason\cite{Lieb,Lieb2}
as consequences of the total order. Indeed all but one  
can be derived, but currently we cannot rule out the existence 
of catalysis\cite{JonathanP99} in the resource theory we 
devised for entanglement: there could be triplet of states 
$\rho, \pi$ and $\sigma$ such that the transformation 
$\rho \rightarrow \pi$ is not possible, but become feasible 
under the presence of the catalyst $\sigma$, i.e. 
$\rho \otimes \sigma \rightarrow \pi \otimes \sigma$. The 
non-existence of catalysis, however, is one of the axioms posed by 
Lieb and Yngvason. Interestingly, it can be readily seen that the existence of 
catalysts is equivalent to the non additivity of the entropy 
function in our theory, given by $E_R^{\infty}$, which is an 
open problem in entanglement theory, with implication 
also to the manipulation of entanglement under local operations 
and classical communication\cite{Brandao3}.   
 
The results of this paper show that 
entanglement theory, and in fact general resource
theories satifying certain properties, have 
a thermodynamical flavor in its structural form. 
Indeed, such a link might be seen as an indication 
of the success of Giles\cite{Giles} and Lieb and Yngvason\cite{Lieb} approach
in identifying the underlying mathematical and logical structure behind
the second law. 

\begin{figure}
  \centering
  \includegraphics[scale=0.4]{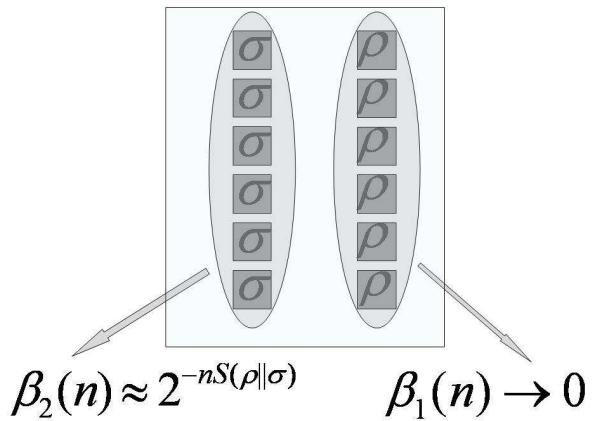}
  \caption{\textbf{Quantum Stein's Lemma.} In quantum hypothesis testing an observer is given several identical copies of an unknown quantum state with the promise that it is described either by the density matrix $\rho$ or $\sigma$. His task is to perform measurements on these copies in order to find out which is the true state of the system. There are two types of errors in the process: $\beta_1(n)$ ($\beta_2(n)$) is the probability that, after performing a measurement on $n$ copies, the observer concludes that the system was described by the state $\sigma$ ($\rho$) when in reality it was described by $\rho$ ($\sigma$). Quantum Stein's Lemma gives the optimal rate of the exponential decay of $\beta_2(n)$ in the limit of infinitely many copies, when one requires that $\beta_1(n) \rightarrow 0$ asymptotically. Its direct part states that any number smaller than $S(\rho || \sigma)$ is an achievable rate. This can be mathematically expressed in a compact manner as $\tr(\rho^{\otimes n} - 2^{yn}\sigma^{\otimes n})_+ \rightarrow 1$ whenever $y < S(\rho || \sigma)$\cite{Petz1}. The converse part, in turn, establishes that there is no test with a better rate than the relative entropy, i.e. $\tr(\rho^{\otimes n} - 2^{yn}\sigma^{\otimes n})_+ \rightarrow 0$ for every $y > S(\rho || \sigma)$\cite{stein}. Here $\tr(A)_+$ stands for the sum of the positive eigenvalues of $A$.}
\end{figure}

\section{Methods}
We present an overview of the main steps taken in 
order to establish eq. (\ref{main}). The argument is rather 
involved and it is therefore out of the scope of this 
paper to present all the details. A rigorous proof 
of theorem I is given in ref. 26. The 
proof can be divided in three parts.  

In the first, we connect $E_R^{\infty}$ to the global 
robustness of entanglement. Let us define
\begin{eqnarray} \label{globalr}
        LG(\rho) := \inf_{\{ \rho_n\in {\cal D}
        ({\cal H}^{\otimes n}): D(\rho_n, \rho^{\otimes n}) 
        \rightarrow 0 \}} \left \{ \lim_{n 
        \rightarrow \infty} \frac{LR_G(\rho_n)}{n} \right \},
\end{eqnarray}
where $LR_G(\rho) := \log(1 + R_G(\rho))$.
This quantity, which we call asymptotic global robustness 
of entanglement, is a proper modification of the global 
robustness of entanglement so that it becomes a meaningful 
quantity in asymptotic considerations. A first step towards 
eq. (\ref{main}) is the proof that for every state $\rho$,
\begin{equation} \label{1111}
        E_R^{\infty}(\rho) = LG(\rho).
\end{equation}
The inequality $E_R^{\infty}(\rho) \leq LG(\rho)$ follows 
from the operator monotonicity of the $\log$ function 
together with the asymptotic continuity of $E_R$\cite{Horodecki5}. 
The converse inequality, in turn, can be derived from the 
converse part of quantum Stein's Lemma, explained in fig. 4. 
Essentially, if $y > S(\rho || \sigma)$, then, for sufficiently
large $n$, we have $\rho_n \leq 2^{y n} 
\sigma^{\otimes n}$, for an approximation $\rho_n$ of $\rho^{\otimes n}$ 
such that $D(\rho^{\otimes n}, \rho_n)\rightarrow 0$ asymptotically. If we take $\sigma$ as a 
separable state such that $E_R(\rho) = S(\rho || \sigma)$, 
we find from the definition of $LR_G$ that
$LR_G(\rho_n)/n \leq E_R(\rho)$, from which follows 
that $LG(\rho) \leq E_R(\rho)$. Finally, we use that 
$LG$ is weakly-additive to obtain eq. (\ref{1111}). 

In the second part, we show that for every state $\rho$,
\begin{equation} \label{2222}
        E_C(\rho) = E_R^{\infty}(\rho).
\end{equation}
This is obtained by showing that $E_C(\rho) = LG(\rho)$, 
which together with eq. (\ref{1111}) implies eq. (\ref{2222}). 
That $E_C(\rho) \geq LG(\rho)$ can be established from the 
fact that $LR_G$ can be increased by a 
$\epsilon$-non-entangling map by no more than $\log(1 + \epsilon)$. 
To prove the converse inequality, we construct a sequence 
of $\epsilon_n$-non-entangling map achieving the lower 
bound $LG(\rho)$. We consider the sequence of maps 
\begin{eqnarray}
        \Psi_n(A) &=& \tr(A (\phi^+)^{\otimes k_n}) \rho_n 
        \nonumber \\ &&+ \tr(A (\id - (\phi^+)^{\otimes k_n}))
        \pi_n,
\end{eqnarray}
where $\{ \rho_n \}$ is an optimal sequence of approximations 
for $\rho^{\otimes n}$ in eq. (\ref{globalr}) and $\pi_n$ an 
optimal state for $\rho_n$ in the global robustness of 
entanglement. By the convexity of $R_G$\cite{Vidal1, Nielsen1} 
and the fact that the maximal overlap of $(\phi^+)^{\otimes k_n}$
with a separable state is $1/2^{k_n}$, we find that by choosing 
$k_n = LR_G(\rho_n)$, each $\Psi_n$ is a 
$R_G(\pi_n)$-non-entangling map. Then, we use that 
(i) $R_G(\rho_n) \rightarrow \infty$ and (ii) $R_G(\pi_n) \leq 1 / R_G(\rho_n)$ 
to see that $\epsilon_n = R_G(\pi_n) \rightarrow 0$ asymptotically. Fact (ii) follows 
directly from the definiion of $R_G$, while fact (i) is a consequence of the quantum 
de Finetti theorem\cite{Renner}. We thus find $\{ \Psi_n \}$ as an admissible sequence of 
maps for $E_C$, from which follows that indeed 
$E_C(\rho) \geq LG(\rho)$.

In the third and last part we show that 
$E_D(\rho) = E_R^{\infty}(\rho)$, completing the proof. That 
$E_D(\rho) \leq E_R^{\infty}(\rho)$ can be established from 
the asymptotic continuity of the relative 
entropy of entanglement\cite{Horodecki5} and the fact that also $E_R$ can 
be increased by a $\epsilon$-non-entangling map by no more than 
$\log(1 + \epsilon)$. 

To show the converse inequality, we proceed in two steps. The first is to derive a more handy expression 
for the singlet-fraction under non-entangling maps, defined 
as $F_{sep}(\rho, 2^r) := \max_{\Psi} \tr(\Psi(\rho) 
(\phi^+)^{\otimes r})$, where the maximum is taken over 
all non-entangling operations $\Psi$. By using arguments from 
convex optimization and duality, it can be shown, in analogy 
with what was done for PPT maps in ref. 27, that
\begin{equation} \label{sfsep}
F_{sep}(\rho^{\otimes n}; 2^{nD}) = \min_{\sigma \in {\cal S}, b \in \mathbb{R}}  \tr (\rho^{\otimes n} - 2^{r n}\sigma)_+ + 2^{(b - D)n},
\end{equation}
where ${\cal S}$ is the set of separable states. If $D$ is such that $\lim_{n \rightarrow \infty} F_{sep}(\rho^{\otimes n}; D) = 1$, then it is clear that $E_D(\rho) \geq D$. From the direct part of quantum Stein's Lemma we have that $\tr(\rho^{\otimes n} - 2^{y n} \sigma^{\otimes n})_+ \rightarrow 1$ whenever $y < S(\rho || \sigma)$ (see fig. 4). The key point of the proof is that a very similar situation holds for  $\min_{\sigma_n \in {\cal S}} \tr (\rho^{\otimes n} - 2^{r n}\sigma_n)_+$, if we replace $S(\rho || \sigma)$ by $E_R^{\infty}(\rho)$. Using the exponential de Finetti theorem recently proved by Renner\cite{Renner}, we show that
\begin{equation} \label{sup}
\lim_{n \rightarrow \infty}  \min_{\sigma_n \in {\cal S}} \tr( \rho^{\otimes n} - 2^{r n}\sigma_n )_+ = 1
\end{equation}
for $r < E_R^{\infty}(\rho)$, which combined with eq. (\ref{sfsep}) readily implies that $E_D(\rho) \geq E_R^{\infty}(\rho)$. \\

{\em Correspondence --} Correspondence and requests for materials
should be addressed to Fernando Brand\~ao.
(Email: {\tt fernando.brandao@imperial.ac.uk})
 
{\em Acknowledgements --}
We gratefully thank K. Audenaert, J. Eisert, A. Grudka, M. Horodecki, R. Horodecki, S. Virmani, and R.F. Werner for 
useful discussions and correspondence. This work is part of the QIP-IRC supproted by EPSRC and the 
Integrated Project Qubit Applications (QAP) supported by the 
IST diectorate and was supported by the Brazilian agency Conselho Nacional 
de Desenvolvimento Cient{\'i}fico e Tecnol{\'o}gico (CNPq) 
and a Royal Society Society Wolfson Research Merit Award.\\

{\em Competing Interests --} The authors declare that they 
have no competing financial interests.\\


\begin{thebibliography}{99}

\bibitem{Thermo} Callen, H.B. Thermodynamics and an Introduction to Thermostatistics, John Wiley and Sons (1985).

\bibitem{Giles} Giles, R. Mathematical Foundantions of Thermodynamics, Pergamon, Oxford, 1964. 

\bibitem{Lieb} Lieb, E.H. and Yngvason, J. The Physics and Mathematics of the Second Law of Thermodynamics. \textit{Phys. Rept.} \textbf{310}, 1-96 (1999).  

\bibitem{Lieb2} Lieb, E.H. and Yngvason, J. The Mathematical Structure of the Second Law of Thermodynamics. \textit{Current Developments in Mathematics, 2001} \textbf{89}, International Press (2002). 

\bibitem{Plenio1} Plenio, M.B. and Virmani, S. An introduction to entanglement measures. \textit{Quant. Inf. Comp.} \textbf{7}, 1-51 (2007).  

\bibitem{Horodecki0} Horodecki, R. Horodecki, P. Horodecki, M. and Horodecki, K. Quantum entanglement. \textit{quant-ph/0702225}. 
 
\bibitem{Bennett00} Bennett, C.H. and Divincenzo D.P. Quantum Information and Computation. \textit{Nature} \textbf{404}, 247-255 (2000).

\bibitem{Einstein} Einstein, A. Podolsky, B. and Rosen, N. Can quantum mechanical description of physical reality be considered complete? \textit{Phys. Rev.} \textbf{47}, 777-780 (1935).

\bibitem{Schrod} Schr\"odinger, E. Die gegenw\"artige Situation in der Quantenmechanik. \textit{Naturwissenschaften} \textbf{23}, 807-812 (1935).

\bibitem{Bennett BCJPW 92} Bennett, C.H. Brassard, G. Crepeau, C. Jozsa, R. Peres, A. Wootters, W.K. Teleporting an unknown quantum state via dual 
classical and Einstein-Podolsky-Rosen channels. \textit{Phys. Rev. Lett.} {\bf 70}, 1895-1899 (1992).

\bibitem{Bennett1} Bennett, C.H. Bernstein, H.J. Popescu, S. and Schumacher, B. Concentrating partial entanglement by local operations. \textit{Phys. Rev. A} \textbf{53}, 2046-2052 (1996). 

\bibitem{Horodecki1} Horodecki, M. Horodecki, P. and Horodecki, R. Mixed-State Entanglement and Distillation: Is there a "Bound" Entanglement in Nature? \textit{Phys. Rev. Lett.} \textbf{80}, 5239-5242 (1998).  

\bibitem{Vidal0} Vidal, G. and Cirac, J.I. Irreversibility in Asymptotic Manipulations of Entanglement. \textit{Phys. Rev. Lett.} \textbf{86}, 5803-5806 (2001).  

\bibitem{Horodecki22} Yang, D. Horodecki, M. Horodecki, R. and Synak-Radtke, B. Irreversibility for all bound entangled states. \textit{Phys. Rev. Lett}. \textbf{95}, 190501-190504 (2005).   

\bibitem{Popescu1} Popescu, S. and Rohrlich, D. Thermodynamics and the Measure of Entanglement. \textit{Phys. Rev. A} 
\textbf{56}, R3319-R3321 (1997).

\bibitem{Horodecki4} Horodecki, P. Horodecki, R. and Horodecki, M. Entanglement and thermodynamical analogies. \textit{Acta Phys. Slov.} \textbf{48}, 141-156 (1998).

\bibitem{Plenio2} Plenio, M.B. and Vedral, V. Entanglement in Quantum Information Theory. \textit{Contemp. Phys.} \textbf{39}, 431-466 (1998).   

\bibitem{Plenio4} Vedral, V. and Plenio, M.B. Entanglement Measures and Purification Procedures. \textit{Phys. Rev. A} \textbf{57}, 1619-1633 (1998). 

\bibitem{HOH02} Horodecki, M. Oppenheim, J. and Horodecki, R. Are the laws of entanglement theory thermodynamical? \textit{Phys. Rev. Lett.} \textbf{89} 240403 (2002).  

\bibitem{Werner1} Werner, R.F. Quantum states with Einstein-Podolsky-Rosen correlations admitting a hidden-variable model. \textit{Phys. Rev. A} \textbf{40}, 4277-4281 (1989). 

\bibitem{Petz1} Hiai, F. and Petz, D. The Proper Formula for Relative Entropy and its Asymptotics in Quantum Probability. \textit{Commun. Math. Phys.} \textbf{143}, 99-114 (1991). 

\bibitem{Renner} Renner, R. Symmetry of large physical systems implies independence of subsystems. \textit{Nature Physics} \textbf{3}, 645-649 (2007). 

\bibitem{stein} Ogawa, T. and Nagaoka, H. Strong Converse and Stein's Lemma in the Quantum Hypothesis Testing. \textit{IEEE Trans. Inf. Theo.} {\bf 46}, 2428-2433 (2000).

\bibitem{Plenio3} Vedral, V. Plenio, M.B. Rippin, M.A. and Knight, P.L. Quantifying entanglement. \textit{Phys. Rev. Lett.} \textbf{78}, 2275-2279 (1997).

\bibitem{Vidal1} Vidal, G. and Tarrach, R. Robustness of entanglement. \textit{Phys. Rev. A} \textbf{59}, 141-155 (1999).  

\bibitem{Nielsen1} Harrow, A.W. and Nielsen, M.A. How robust is a quantum gate in the presence of noise? \textit{Phys. Rev. A} \textbf{68}, 012308-012321 (2003).   

\bibitem{BrandaoPlenio} Brand\~ao, F.G.S.L. and Plenio, M.B. Reversibility of entanglement manipulation under non-entangling operations. \textit{arXiv:0710.5827 [quant-ph].}

\bibitem{Rains1} Rains, E.M. A semidefinite program for distillable entanglement. \textit{IEEE T. Inform. Theory} \textbf{47}, 2921-2933 (2001).

\bibitem{Horodecki5} Donald, M.J. and Horodecki, M. Continuity of Relative Entropy of Entanglement. \textit{Phys. Lett. A} \textbf{264}, 257-260 (1999). 

%\bibitem{Vollbrecht} Vollbrecht, K.G.H. and Werner R.F. Entanglement Measures under Symmetry. \textit{Phys. Rev. A} \textbf{64}, 062307 (2001).  

\bibitem{GourS07} Gour, G. and Spekkens, R.W. The resource
theory of quantum reference frames. \textit{arXiv:07110043 [quant-ph]}.
%
\bibitem{maurerwolf} Maurer, U. Secret key agreement by public discussion from common information. \textit{IEEE Trans. Inf. Theo.} \textbf{39}, 733-742 (1993).

\bibitem{EisertP03} Eisert, J. and Plenio, M.B. Introduction to the basics of entanglement theory in continuous-variable systems. \textit{Int. J. Quant. Inf.} {\bf 1}, 479-506 (2003).
%
\bibitem{JonathanP99} Jonathan, D. and Plenio, M.B. Entanglement-assisted local manipulation of pure quantum states. \textit{Phys. Rev. Lett.} {\bf 83}, 3566-3569 (1999).

\bibitem{Brandao3} Brand\~ao, F.G.S.L. Horodecki, M. Plenio, M.B. and Virmani, S. Remarks on the equivalence of full additivity and 
monotonicity for the entanglement cost. \textit{Open System \& Information Dynamics} \textbf{14}, 333-339 (2007).

\end{thebibliography}
\end{document}